\documentclass[11pt,twoside]{article}
\usepackage{./asp2014}

\aspSuppressVolSlug
\resetcounters

\begin{document}

\title{Reaching Diverse Groups in Long-Term Astronomy Public Engagement Efforts}
\author{Melanie Archipley,$^{1,5}$ Hannah S. Dalgleish,$^{2,5}$ Eva-Maria Ahrer,$^{3,5}$ Daniel Mortimer$^{4,5}$
\affil{$^1$Dept. of Astronomy, University of Illinois Urbana-Champaign, 1002 West Green Street, Urbana, IL, 61801, USA; \email{melanie@iayc.org}}
\affil{$^2$Dept. of Physics, University of Namibia, Pionierspark, Windhoek, Namibia;}
\affil{$^3$Dept. of Physics, University of Warwick, Gibbet Hill Road, Coventry, CV4 7AL, UK;}
\affil{$^4$Cavendish Laboratory, University of Cambridge, 19 J.J. Thomson Avenue, Cambridge, CB3 0HE, UK;}
\affil{$^5$The International Workshop for Astronomy e.V., Türkenstraße 12, Garching, Munich, 85748, DE}}

\paperauthor{Melanie Archipley}{melanie@iayc.org}{0000-0002-0517-9842}{University of Illinois Urbana-Champaign}{Dept. of Astronomy}{Urbana}{IL}{61801}{USA}
\paperauthor{Hannah S. Dalgleish}{hannah@iayc.org}{0000-0002-8970-3065}{University of Namibia}{Dept. of Physics}{Pionierspark}{Windhoek}{Khomas}{Namibia}
\paperauthor{Eva-Maria Ahrer}{eva-maria@iayc.org}{0000-0003-0973-8426}{University of Warwick}{Dept. of Physics}{Coventry}{}{CV4 7AL}{UK}
\paperauthor{Daniel Mortimer}{daniel@iayc.org}{0000-0003-0067-0437}{University of Cambridge}{Cavendish Laboratory}{Cambridge}{}{CB3 0HE}{UK}

\begin{abstract}
Professional astronomy is historically not an environment of diverse identities. In recognizing that public outreach efforts affect career outcomes for young people, it is important to assess the demographics of those being reached and continually consider strategies for successfully engaging underrepresented groups. One such outreach event, the International Astronomical Youth Camp (IAYC), has a 50-year history and has reached $\sim$1700 participants from around the world. We find that the IAYC is doing well in terms of gender (59\% female, 4.7\% non-binary at the most recent camp) and LGBT+ representation, whereas black and ethnic minorities are lacking. In this proceeding, we report the current landscape of demographics applying to and attending the IAYC; the efforts we are making to increase diversity amongst participants; the challenges we face; and our future plans to bridge these gaps, not only for the benefit of the camp but for society overall.
\end{abstract}

\section{Introduction}
Motivating young people to pursue careers in science has become the mission of many outreach programs today. Several studies have shown that informal learning environments are key drivers behind improving attitudes towards science (e.g. \citealt{vennix2017,kitchen2018,roberts2018}). Looking at summer science camps in particular, the International Astronomical Youth Camp (IAYC) is a 50-year old event which uses astronomy as a medium for scientific inquiry. The first camp took place in 1969 in West Germany and the most recent in 2019 was also in Germany --- no in-person 2020 iteration took place due to the worldwide COVID-19 pandemic. There have been 15 different host countries throughout Europe, northern Africa, and the Middle East in the IAYC's history. In 1979, The International Workshop for Astronomy e.V. (IWA) was founded as the official organization behind the camp.

Once a year, around 65 participants aged 16-24 spend three weeks working on an independent project of their choosing. They usually work in pairs or small groups with the guidance of their group leader. There are no lectures or exams. At the end of the camp, the participants write up their findings in a 4-8 page report using \LaTeX, which is published in a report book distributed to all attendees of that year's camp.

Participants come from a great range of backgrounds --- academically and culturally --- and $\sim$80\% have never studied astronomy before attending.\footnote{The IAYC survey comprised of 54 questions and was completed by 307 former participants between 16 August 2017 and 21 September 2019. The results were compiled by IWA and many of the survey’s findings are published in \cite{2018capc.conf..206D,2019NatAs...3.1043D}.} The application process reflects this expectation: applicants do not submit transcripts, grades, a resume, or letters of recommendation. Applicants are prompted to discuss their experiences and inspirations in science and international environments in a ``letter of motivation.'' Since 2018, particular effort has been made to better understand our demographics and how they have changed over the IAYC's long history. Using self-reported data from the IAYC survey,$^1$ internal (generalized and anonymized) data, and anecdotal experience as organizers of the camp, we have identified specific areas of strength (gender, LGBT+) and weakness (racial and ethnic diversity) in terms of diversity of the camp.

In Section \ref{sec:quant}, we present data describing demographics of IAYC participants and leaders from 1969 to 2019. In Section \ref{sec:improve}, we discuss recent efforts to improve our reach to, and welcoming of, underrepresented groups and to what end those efforts have been successful. Finally, in Section \ref{sec:reflect} we assess further gaps to be filled, such as increased funding for low-income participants, to help us achieve our goals in the future.

\section{Quantitative History and Status of Diversity} \label{sec:quant}
For our assessments, we identify four broad categories of diversity in which we are interested: gender, LGBT+, nationality, and ethnicity. We later plan to extend our studies to assess our reach and inclusion of neurodiverse people; people with disabilities; people from varying socioeconomic backgrounds; and more.

\subsection{Gender and LGBT+}
Our earliest records on gender comes from IAYC 1984, with data for every year since then except 1995. Figure \ref{fig:gender_years} shows the gender representation for these 33 camps for participants, alongside 24 years of data for leaders, with the notable exception that an openly non-binary individual attended the camp in 2017 for the first time.

\articlefigure[width=1\textwidth]{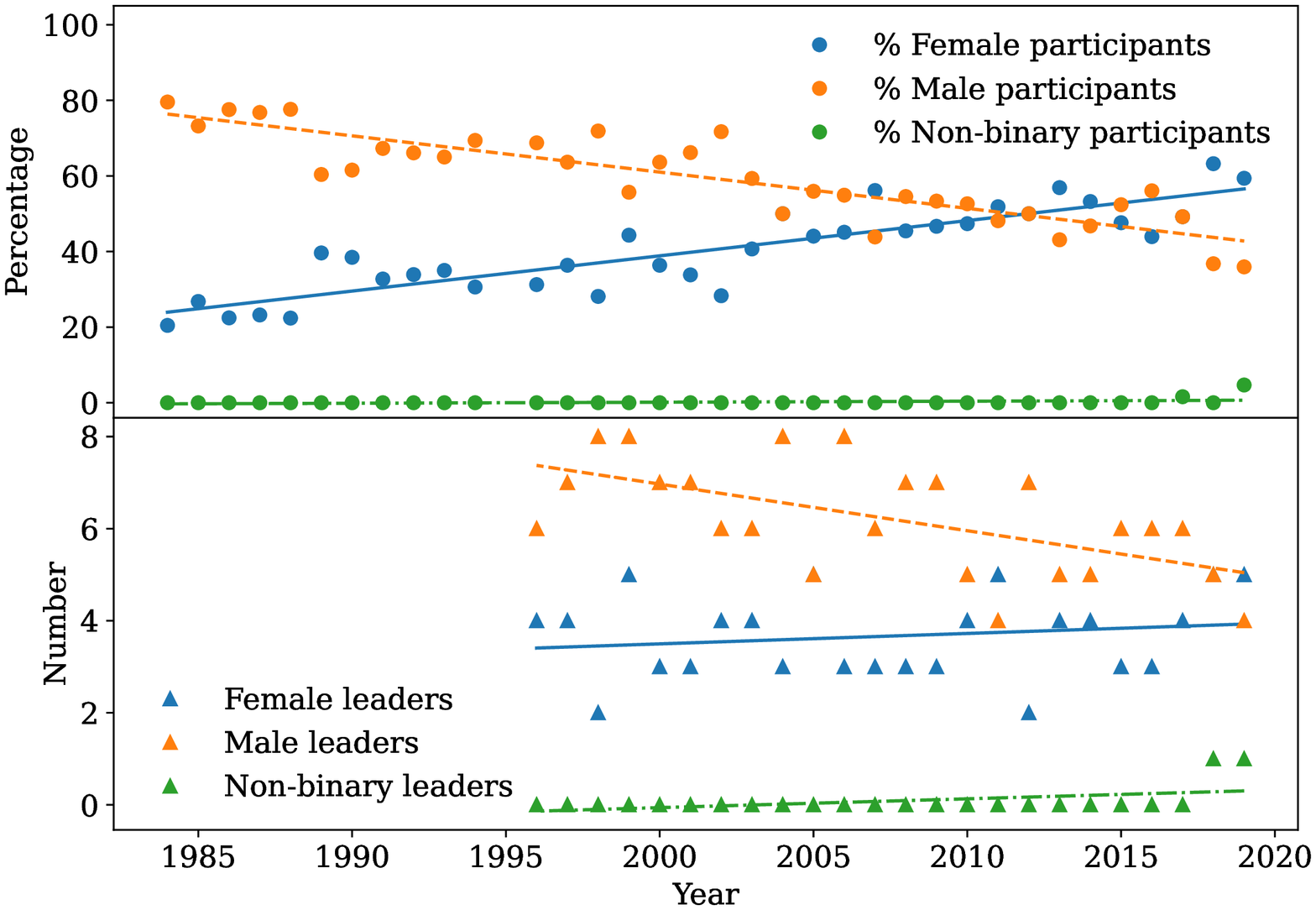}{fig:gender_years}{Data showing the trend of gender representation at the IAYC since 1984. Orange, blue, and green points represent male, female, and non-binary people, respectively. Each straight line is the linear best fit to the data, showing the overall trend of increasing females and decreasing males, even beyond gender parity for participants \textit{(top panel)}. The leader data \textit{(bottom panel)} is represented as raw numbers instead of percentages due to the small size of leaders per camp. From this, we note that the trend of fewer male leaders at the camp over the years is due to a marginal increase in female leaders and a decrease in leader team size overall.}

We infer that the gender gap trended towards parity for most of the IAYC's history, and the gap was finally closed for participants around 2012. Somewhat surprisingly, the gender representation has not stopped trending towards increasing females and decreasing males in the past decade. It is possible that the trend towards gender balance on the leader team had a ripple effect to participants (bottom panel of Figure \ref{fig:gender_years}), however, the leader team has not seen as drastic an increase in female representation.

We would be remiss to not also note that a wide range of gender identities are absent in our limited and statistical discussion of gender. We strive to make the camp atmosphere as open and welcoming as possible to people beyond these limits --- see Section \ref{sec:improve}.
Given the high number of nationalities at each camp, we have a unique opportunity to examine LGBT+ perceptions across many cultures. We are interested in the impact of the camp on LGBT+ and non-LGBT+ participants, as individuals and as youth astronomers and scientists. The beginnings of these efforts exist in the form of a qualitative survey with quotes and feedback presented in \cite{miller_nikki_2021_4554514}. A quantitative survey is now being developed to better assess the LGBT+ population at the IAYC. Overall, we feel that our current representations of gender and LGBT+ people indicate that the IAYC is a welcoming and accessible environment to these diverse groups.

\subsection{Nationality and Ethnicity}
The number of nationalities at each IAYC has steadily and significantly increased since its beginnings (see Figure \ref{fig:nat_years}). However, unique ethnicities have not been represented equitably, and the majority of participants have been European whites. Of the 307 respondents to the IAYC survey (see Footnote 1), 84.7\% were white and not a single respondent was Black (other responses included Mixed, 5.9\%; Asian, 4.2\%; Other, 3.3\%; and Arab, 2.0\%). Participants are aware of, and concerned about, the lack of ethnic diversity at the camp, as the topic has been raised to IWA in recent years.

\articlefigure[width=1\textwidth]{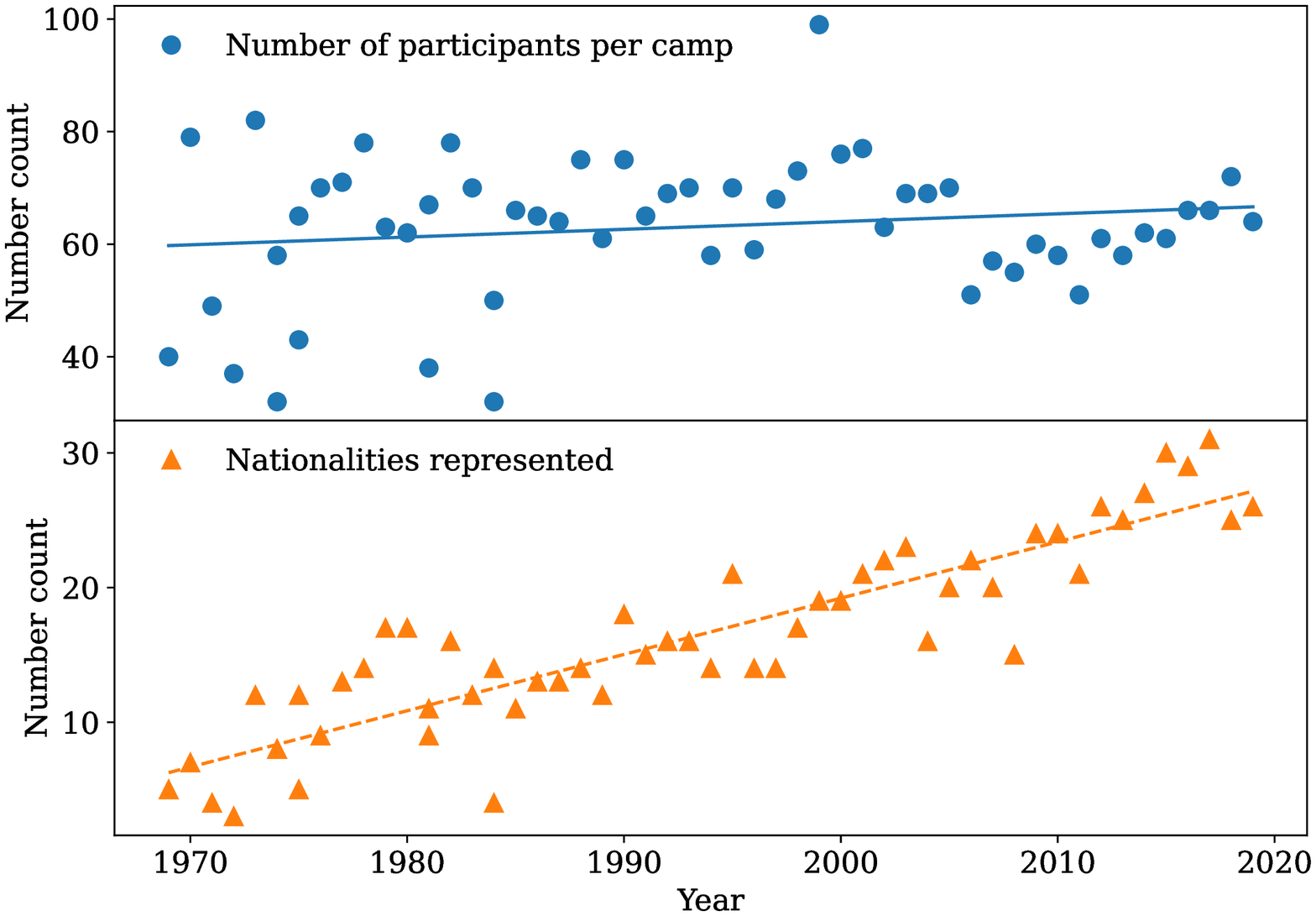}{fig:nat_years}{Data showing the number of nationalities represented at every IAYC since its inception, including the four years that two camps took place. The straight line is the best linear fit to the data. Roughly every decade, five more nationalities are represented at the camp, without appreciably increasing the total number of participants accepted to each camp.}

Even though our demographic is largely European --- and predictably so, given that the last time the camp was outside of Europe was in 1984 --- the lack of Black and minority ethnic (BME) participants is neither representative of the European population as a whole, nor of the professional astronomy community.\footnote{While ethnicity data are hard to come by, we are aware of this demographic survey carried out by the Royal Astronomical Society \citep{ras2017} which gives some indication.} Because of our success in increasing number of nationalities over the years, we know it is possible to extend the IAYC's reach to new countries, but we are nonetheless struggling to reach non-whites. Some of the ways we are strategizing to reach ethnically diverse groups are described in Section \ref{sec:improve}.

\section{Recent Developments to Improve Access} \label{sec:improve}
We recognize the importance of being efficient and prudent in how we tackle the many steps that a participant takes to come to the IAYC. Therefore, we have made adjustments both in the recruitment and application phases as well as during the camp itself. Here we highlight some of the ways we have approached increasing and welcoming diversity.

\subsection{Application Evaluation}
We are not alone in exploring different methods to improve inclusivity through our application process \citep{Johnson_2020,2020arXiv200613685A}. In our initial review process, we experimented with blinding identifying and demographic information --- name, nationality, gender, and age --- from the evaluators with the goal of reducing implicit bias. It is difficult to assess if we achieved that goal explicitly, but some evaluators reported that they felt they were not getting the full picture of applicants without those details. We are exploring ways to fill in those gaps without exposing ourselves to pitfalls of internally-held stereotypes.

After accepting a first round of applicants, we create a waiting list and accept new people as people drop out. Inspired by the Entrofy algorithm for candidate selection \citep{2019arXiv190503314H}, IWA developed an algorithm that assessed the waiting list for an underrepresented nationality each time a spot opened up if the next few candidates were nearly equally ranked. An example is as follows:

\begin{itemize}
    \item[] \textit{Someone declines their acceptance. The next person on the waiting list is from country A. Many confirmed participants are also from country A. The person after them is from country B. There are 0 confirmed participants from country B. The algorithm recommends filling the open spot with the person from country B.}
\end{itemize}

The algorithm was not used on the first round of acceptances and was not a replacement for meritocratic evaluation of applications. In the end, the algorithm was used so rarely and conservatively that no participant who attended the camp was affected by it. Though we are unlikely to employ it in future rounds of evaluation, developing it sparked important discussion and ideas about our application process through this lens.

\subsection{LGBT+ Focused Efforts}
It would not be enough to stop at the recruitment and application stage --- therefore we take deliberate actions to welcome diverse groups once at the camp. In 2017, leaders began to introduce themselves alongside their pronouns and encouraged participants to do the same. In 2019, we added an ``other (if gender is other, please specify)'' option on the application for the ``gender'' category as well as providing the possibility for mixed gender dorm rooms, which the majority of participants chose. These changes were designed specifically to remove any pressure upon participants to slot themselves into a gender identity with which they did not feel wholly comfortable.

The camp has been rooted in celebrations of cultural and ethnic diversity since it originated, with specific time carved out to share and exchange food, music, dance, language, and history across nations. Since 2017, many participants and leaders have taken part in a ``IAYC pride'' celebration during the camp, as an added dimension to the kinds of exchange already taking place. Some annual events include a short informational talk about sexual orientation and gender, a panel discussion, collaborative art and music, and lessons on historical events and people related to the struggle for equity and recognition in the LGBT+ community.

\section{Reflections and Future Plans} \label{sec:reflect}
Addressing diversity in STEM outreach is a continuous effort of self-evaluation, trial and error, data collection and analysis, and sometimes, fundraising. The IAYC has been oversubscribed only in the past $\sim7$ years, significantly so for the two most recent camps which were oversubscribed by a factor of $\sim$2.5. This recent oversubscription calls upon the organizers to ensure that people of all identities and backgrounds are able to attend the camp going forward. Thinking critically about the entire process from the participant's perspective --- deciding to apply, writing a successful application, traveling to the camp, thriving while there, and finally, embracing lessons learned and the IAYC community after the camp --- is crucial to developing equitable outcomes.

With little effort, we have achieved gender parity at the IAYC, as well as a significant upward trend in number of unique nationalities at each camp. LGBT+ identities are now not only well represented but also celebrated by all during the camp. However, we have a poor record when it comes to ethnicity and serving non-white participants. Part of addressing this gap requires financial means we do not currently possess, such as to fund participants' travel from countries outside of Europe. While the actions we described in Section \ref{sec:improve} were focused on aspects like implicit bias and accommodation, we recognize that some solutions will come from simple fundraising.

For the 2020 camp (cancelled due to COVID-19), we included a new optional diversity monitoring form. Questions included parents' level of education, physical or learning disabilities, and sexual orientation in addition to the standard questions about age, gender, and so on. This is a first step to assessing the pipeline from application to acceptance and participation. We intend to monitor these data going forward to inform our choices when organizing the camp to ensure its accessibility to all.

\acknowledgements MA acknowledges support from the Illinois Graduate College's Conference Presentation Award. HD acknowledges support from the UKRI STFC Global Challenges Research Fund project ST/S002952/1. Figures are produced using the python package \texttt{matplotlib} \citep{Hunter:2007}.

\bibliographystyle{aasjournal}
\bibliography{aspbib}{}  

\end{document}